# The structure and magnetism of graphone

*L. Feng, W. X. Zhang\**

Computational Condensed Matter Physics Laboratory, Department of Physics, Taiyuan University of Technology, Taiyuan 030024, People's Republic of China

Graphone is a half-hydrogenated graphene. The structure of graphone is illustrated as trigonal adsorption of hydrogen atoms on graphene at first. However, we found the trigonal adsorption is unstable. We present an illustration in detail to explain how a trigonal adsorption geometry evolves into a rectangular adsorption geometry. We check the change of magnetism during the evolution of geometry by evaluating the spin polarization of the intermediate geometries. We prove and clarify that the rectangular adsorption of hydrogen atoms on graphene is the most stable geometry of graphone and graphone is actually antiferromagnetic.

Graphene, a single atomic layer of carbon atoms, is a good candidate for nano-electronic devices of atomic thickness.[1-3] With band crossing at the Dirac points of the Brillouin zone, the pristine graphene is a semimetal and its conducting properties are remarkable.[4,5] In order to apply graphene into electronics and photonics,[6] a great interest on graphene functionalization is stimulated. Chemical functionalization[7-11] is a useful tool to tune the band structure and the majority carrier type of graphene for applications.[12-17] It has been experimentally demonstrated that mobility of charge carriers can be tuned in highly conducting semimetallic graphene with chemisorption of atomic hydrogen.[7] M. Z. S. Flores et al. suggest that large domains of perfect graphane-like structures are unlikely to be formed.[18] J. Zhou et al. investigate semihydrogenated graphene which is named as graphone with the configuration shown in Fig. 1.[19] Their calculations indicate that the unhydrogenated carbon atoms couple ferromagnetically. Based on the geometry of Fig. 1(a) (We name it as triangular graphone), using geometry optimization we find the stable structure is the geometry shown in Fig. 1(c), and we name it as rectangular graphone. We have made geometry optimizations based on three geometries different from the triangular one. We found that the rectangular graphone is still the most stable, i.e. having the lowest energy -1300.8462 eV. We have shown one example in Fig. 1(b). The electronic structure calculation indicates the rectangular graphone is antiferromagnetic and has an indirect band gap of ～2.45eV, which is very different from the triangular graphone that is ferromagnetic and has a smaller band gap of ～0.67eV.

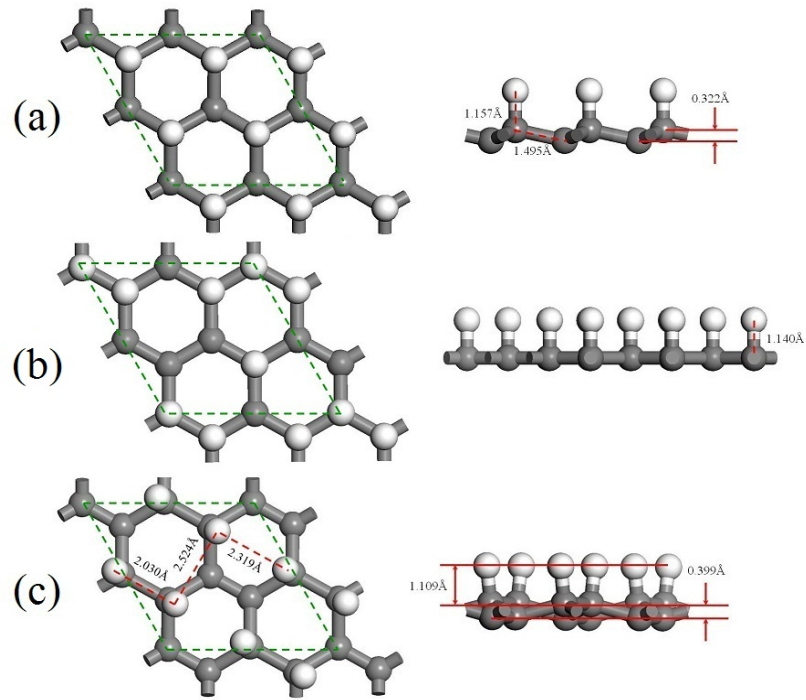

Figure 1. (color online) (a) and (b) are the initial geometries of graphone before optimization. (c) is the optimized geometry of graphone. Gray ball is carbon atom. White ball is hydrogen atom. Left side is top view with the green rhombus showing the unit cell. Right side is side view showing C-C and C-H distances.

The spin-polarized density functional theory (DFT) calculations are performed with CASTEP code[20] using the generalized gradient approximation (GGA), Perdew-Burke-Ernzerholf (PBE) exchange correlation functional and ultrasoft pseudopotential. In geometry optimization we use triangular graphone and three other different geometries as initial ones. To avoid interactions between two sheets, vacuum space is set as 15 Å. The k-points in reciprocal space are set as $7\times 7\times 1$, following the Monkhorst–Pack scheme. The structure is relaxed with a cutoff energy of 310eV. No symmetry constraints are used. The convergence of energy and force are set as $10^{-4}$ eV and 0.01 eV/Å, respectively. The geometry in Reference 19 is shown in Fig. 1(a). The bond lengths of C-C and C-H are 1.495 Å and 1.157 Å, respectively. The distance between two carbon planes is 0.322 Å.

The most stable geometry is shown in Fig. 1(c). The bond length of C-H is 1.133 Å, smaller than that of triangular geometry. Different from triangular geometry, the bond of C-H is not perpendicular to carbon planes. All the hydrogen atoms are still in the same plane. The distances between the hydrogen atoms are $d_{12}$=2.030 Å, $d_{23}$=2.524 Å and $d_{34}$=2.319 Å, respectively, as shown in Fig. 1(c). The distance between two carbon planes is 0.399 Å. This distance is a little larger than that of triangular graphone. The distance between top carbon plane and hydrogen plane is 1.109 Å. Hornekaer L. et al. has investigated the topograph of hydrogenated graphene by STM.[21] In the STM image of Reference 21 they found dimmer of hydrogen atoms is the most favorable adsorption geometry which is consistent with our calculation. And there is no adsorption corresponding to triangular graphone. This is a direct proof for the conclusion that the rectangular adsorption is the most stable geometry.

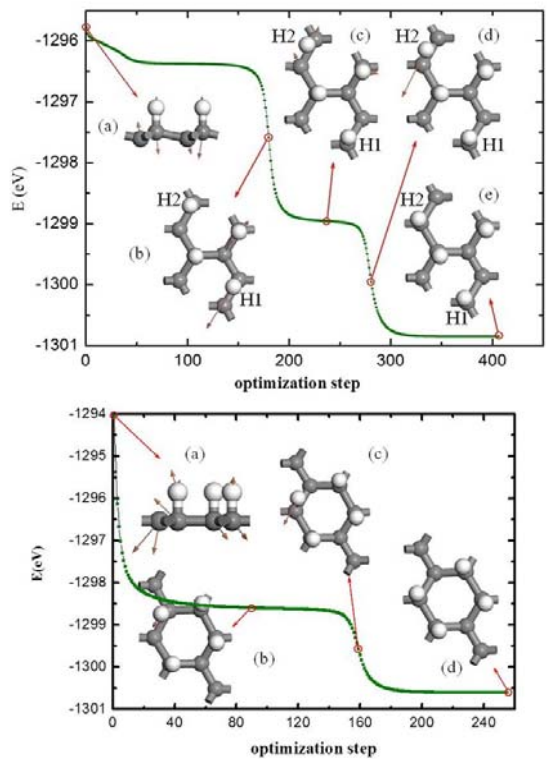

Figure 2. (color online) Dependence of energy on optimization step. The top panel is corresponding to the triangular graphone. The geometries and forces on each atom are shown for the corresponding step: (a) n=1 (triangular graphone), (b) n=175, (c) n=240, (d) n=278 and (e) n=417 (rectangular graphone). The bottom panel is corresponding to the geometry shown in Fig. 1(b). The geometries and forces on each atom are shown for the corresponding step: (a) n=1, (b) n=90, (c) n=160, and (d) n=260 (rectangular graphone).

For triangular graphone , the dependence of energy on optimization step is shown in the top panel of Fig. 2. The evolution of geometry takes 417 steps. The insets show the geometry and the forces on each atom for corresponding step, n=1, 175, 278 and 417. The inset (a) shows that the forces on carbon and hydrogen atoms of triangular graphone are nearly perpendicular to the graphene plane. The inset (b) shows the intermediate geometry corresponding to the first step of energy decrease. Two hydrogen atoms (H1, H2 in the insets) deviate from their original sites. The force on the H1 atom is considerable. It can be concluded that the triangular graphone is unstable. That is, once a perturbation along the graphene plane appears in triangular graphone, the hydrogen atoms tend to move away from their original sites. The total energy decreases dramatically due to the hydrogen atom moving to the top site above the carbon atom which the force directs to. After the jump of H1 atom, the geometry reaches to another unstable state and the energy shows a platform. This unstable geometry is depicted in the inset (c). The H1 atom arrives at a stable position and the H2 atom deviates a little from its original position. There are no remarkable forces on hydrogen atoms. The forces on the carbon atoms are small and parallel to the graphene plane. The inset (d) shows the geometry corresponding to the second step of energy decrease. There is a large force on the H2 atom. It can be deduced that the H2 atom moves to the top site above carbon atom which the force directs to. After the jump of H2 atom, the geometry reaches to the stable state, that is, the forces on each atom are almost zero, as shown in the inset (e). For the initial geometry in Fig. 1(b), the energy dependence on optimization step is shown in the bottom panel in Fig. 2. The evolution is similar to the triangular one, and the final stable geometry is also the rectangular geometry.

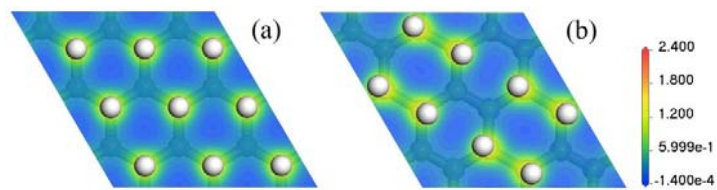

Figure. 3 (color online) Calculated electron densities of (a) triangular graphone and (b) rectangular graphone. Both of the slices are located in the very middle of top carbon plane and hydrogen plane. The unit is e/Å$^3$.

To understand the change of electronic interaction, electronic densities for the triangular graphone and rectangular graphone are plotted in Fig. 3. Both of the slices in Fig. 3 are located in the very middle of the top carbon plane and hydrogen plane. In triangular graphone the distribution of electrons are localized around the hydrogen atoms, and there are almost no electrons distributed between two nearest neighbor hydrogen atoms, where the interaction is very weak. However, in rectangular graphone the electrons are not localized around the hydrogen atoms, and the electron density between two nearest neighbor hydrogen atoms is considerable, so the interaction between two nearest neighbor hydrogen atoms is enhanced in rectangular graphone, and the geometry is more stable than triangular graphone.

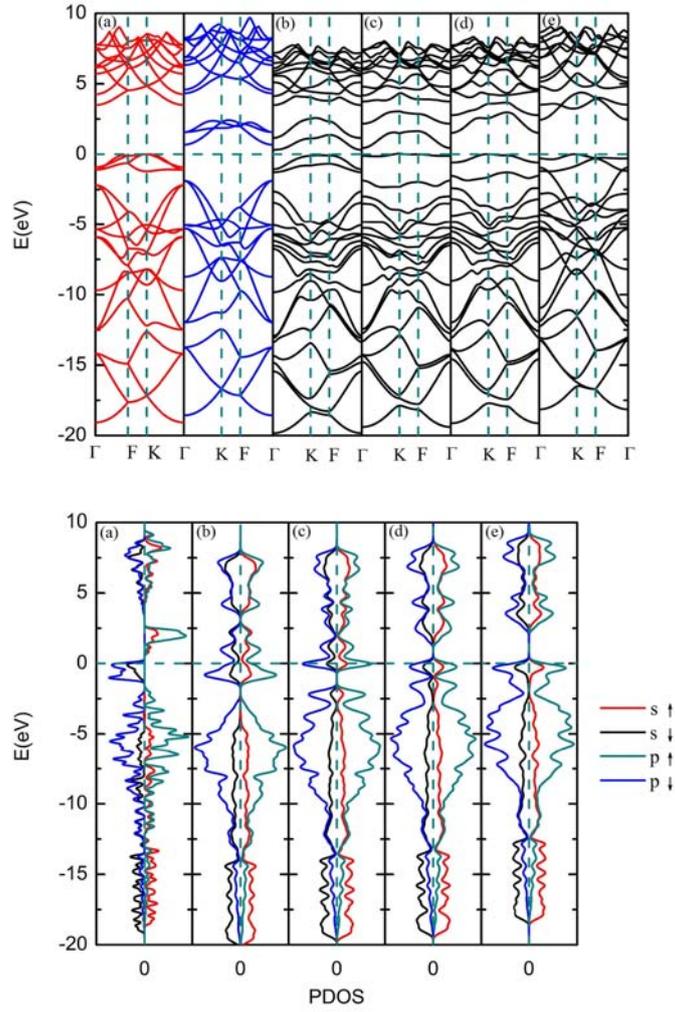

Figure. 4 (color online) Calculated band structures (top panel) and PDOS (bottom panel) of the corresponding geometry in the evolution of triangular graphone for (a) n=1 (triangular graphone), (b) n=175, (c) n=240, (d) n=278 and (e) n=417 (rectangular graphone), respectively. In top panel, Γ =(0, 0, 0), K= (-1/3, 2/3, 0) and F= (0, 1/2, 0) are shown. K is the Dirac point of graphene. In bottom panel, black curve represents the spin up s-electrons, red curve represents the spin down s-electrons, blue curve represents the spin up p-electrons, and cyan curve represents the spin down p-electrons.

Since the triangular graphone is not stable and its energy is ~5eV higher than the rectangular geometry, it is impossible to introduce magnetism by constructing this geometry in practice. Then we are interested in whether the rectangular graphone is magnetic. Fig. 4 shows the band structures and calculated projected density of state (PDOS) of the five representative geometries in the evolution of triangular graphone (shown in the insets of the top panel of Fig. 2) Besides the triangular graphone in which a clear magnetic splitting is observed, the net moments for other structures are zero, indicating that the rectangular graphone is antiferromagnetic. In the triangular graphone, each unhydrogenated carbon atom has an unpaired spin, which provides one Bohr mageneton. In the rectangular graphone, the two unhydrogenated carbon atoms have two spins, which are paired and the rectangular graphone is antiferromagnetic. Examination of the states in the vicinity of the Fermi level shows that the two doping states above the Fermi level shift to a higher energy region, resulting in an energy gap of ~2.45eV. Our findings are in accordance with Lieb's theorem.[22]

In conclusion, using density functional theory we investigated the structure and magnetism of graphone. Based on the triangular graphone and three other initial geometries, the geometry optimization has been carried out. It is found the triangular graphone is unstable and the rectangular graphone is the most stable. In the process of optimization of triangular graphone, there are two physical steps corresponding to obvious energy decrease. The two hydrogen atoms move towards the neighboring carbon atoms from the starting position of triganular geometry, and finally form the stable geometry of rectangular graphone. Comparison of the electron density in triangular graphone with that in rectangular graphone indicates that there is interaction between two nearest neighbor hydrogen atoms in rectangular graphone but in triangular graphone，which explains the rectangular graphone is stable and the triangular grpahone is unstable. Our calculation also indicates the rectangular graphone is an antiferromagnetic semiconductor with an indirect band gap of ∼2.45eV, different from the triangular grapone which is a ferromagnetic semiconductor with a smaller indirect band gap of ∼0.67eV. The findings above indicate the stable geometry of graphone is antiferromagnetic.

*To whom correspondence should be addressed: zhangwenxing@tyut.edu.cn

ACKNOWLEDGMENT

This work is supported by the National Natural Science Foundation of China in Grant No.11147142.

REFERENCES

(1) Novoselov, K. S.; Geim, A. K.; Morozov, S. V.; Jiang, D.; Zhang, Y.; Dubonos, S. V.; Grigorieva, I. V.; Firsov, A. A. Science **2004**, 306, 666-669.

(2) Gass, M. H.; Bangert, U.; Bleloch, A. L.; Wang, P.; Nair, R. R.; Geim, A. K. Nat. Nanotech. **2008**, 3, 676-681.

(3) Stankovich, S.; Dikin, D. A.; Dommett, G. H. B.; Kohlhaas, K. M.; Zimney, E. J.; Stach, E. A.; Piner, R. D.; Nguyen, S. T.; Ruoff, R. S. Nature. **2006**, 442, 282-286.

(4) Geim, A. K. Science **2009**, 324, 1530-1534.

(5) Katsnelson, M. I. Mater. Today **2007**, 10, 20-27.

(6) Novoselov, K. Nat. Mater. **2007**, 6, 720-721.

(7) Elias, D. C.; Nair, R. R.; Mohiuddin, T. M. G.; Morozov, S. V.;Blake, P.; Halsall, M. P.; Ferrari, A. C.; Boukhvalov, D. W.; Katsnelson, M. I.; Geim, A. K.; Novoselov, K. S. Science **2009**, 323, 610–613.

(8) Robinson, J. T.; Burgess, J. S.; Junkermeier, C. E.; Badescu, S. C.; Reinecke, T. L.; Perkins, F. K.; Zalalutdinov, M. K.; Baldwin, J. W.; Culbertson, J. C.; Sheehan, P. E.; Snow, E. S. Nano Lett. **2010**, 10, 3001–3005.

(9) Englert, J. M.; Dotzer, C.; Yang, G.; Schmid, M.; Papp, C.; Gottfried, J. M.; Steinruck, H.-P.; Spiecker, E.; Hauke, F.; Hirsch, A. Nat. Chem. **2011**, 3, 279–286.

(10) Sahin, H.; Ataca, C.; Ciraci, S. Appl. Phys. Lett. **2009**, 95, 222510-1–222510-3.

(11) Casolo, S.; Lovvik, O. M.; Martinazzo, R.; Tantardini, G. F. J. Chem. phys. **2009**, 130, 054704-1–054704-10.

(12) Zhang, Y. B.; Tan, Y. W.; Stormer, H. L.; Kim, P. Nature **2005**, 438, 201–204.

(13) Gao, H. L.; Wang, L.; Zhao, J. J.; Ding, F.; Lu, J. P. J. Phys. Chem. C **2011**, 115, 3236–3242.

(14) Haberer, D.; Vyalikh, D. V.; Taioli, S.; Dora, B.; Farjam, M.; Fink, J.; Marchenko, D.; Pichler, T.; Ziegler, K.; Simonucci, S.; Dresselhaus, M. S.; Knupfer M.; Buchner, B.; Gruneis, A. Nano Lett. **2010**, 10, 3360–3366.

(15) Jeon, K.-J.; Lee, Z.; Pollak, E.; Moreschini, L.; Bostwick, A.; Park, C.-M.; Mendelsberg, R.; Radmilovic, V.; Kostecki, R.; Richardson, T. J.; Rotenberg, E. ACS Nano **2011**, 5, 1042–1046.

(16) Cheng, S.-H.; Zou, K.; Okino, F.; Gutierrez, H. R.; Gupta, A.; Shen, N.; Eklund, P. C.; Sofo, J. O.; Zhu, J. Phys. Rev. B **2010**, 81, 205435.

(17) Samarakoon, D. K.; Wang, X. Q. ACS Nano, **2009**, 3, 4017–4022.

(18) Flores, M. Z. S.; Autreto, P. A. S.; Legoas, S. B.; Galvao, D. S. Nanotech. **2009**, 20, 465704-1–465704-6.

(19) Zhou, J.; Wang, Q.; Sun, Q.; Chen, X. S. Kawazoe, Y.; Jena, P. Nano lett. **2009**, 9, 3867–3870.

(20) Clark, S. J.; Segall, M. D.; Pickard, C. J.; Hasnip, P. J.; Probert, M. I. J.; Refson, K.; Payne, M. C. Z. Kristallogr. **2005**, 220, 567-570.

(21) Hornekaer, L.; Sljivancanin, Z.; Xu, W.; Otero, R.; Rauls, E.; Stensgaard, I.; Laegsgaard, E.; Hammer, B.; Besenbacher, F. Phys. Rev. Lett. **2006**, 96, 156104-1–156104-4.

(22) Lieb, E. H.; Phys. Rev. Lett. **1989**, 62, 1201–1204.